\documentclass[preprint,aps,amsmath,amssymb,superscriptaddress,floatfix]{revtex4-2}

\usepackage[utf8]{inputenc}
\usepackage{setspace}
\usepackage{graphicx}
\usepackage{color}
\usepackage{soul}
\usepackage{hyperref}
\usepackage{miller}
\usepackage{pdfpages}
\usepackage{bbold}
\usepackage{gensymb}
\usepackage{xr}

\newcommand{\simlt}
      {\ifmmode       { \raisebox{-.8em}{$<$}\atop\sim}
         \else        {$\raisebox{-.8em}{$<$}\atop\sim$}
      \fi}

\makeatletter
\def\@hangfrom@section#1#2#3{\@hangfrom{#1#2}#3}
\def\@hangfroms@section#1#2{#1#2}
\makeatother

\makeatletter
\AtBeginDocument{\let\LS@rot\@undefined}
\makeatother

\usepackage[version=4]{mhchem}

\begin{document}


\title{Electronic structure of monolayer-\ce{CrTe2}: an antiferromagnetic 2D van der Waals material}

\author{Olivia Armitage}
\affiliation{SUPA, School of Physics and Astronomy, University of St Andrews, North Haugh, St Andrews, KY16 9SS, United Kingdom}
\author{Naina Kushwaha}
\affiliation{SUPA, School of Physics and Astronomy, University of St Andrews, North Haugh, St Andrews, KY16 9SS, United Kingdom}
\author{Akhil Rajan}
\affiliation{SUPA, School of Physics and Astronomy, University of St Andrews, North Haugh, St Andrews, KY16 9SS, United Kingdom}
\author{Luke C. Rhodes}
\author{Sebastian Buchberger}
\author{Bruno Kenichi Saika}
\author{Shu Mo}
\affiliation{SUPA, School of Physics and Astronomy, University of St Andrews, North Haugh, St Andrews, KY16 9SS, United Kingdom}
\author{Matthew D. Watson}
\affiliation{Diamond Light Source Ltd., Harwell Science and Innovation Campus, Didcot OX11 0DE, UK}
\author{Phil D. C. King}
\email[Correspondence to: ]{pdk6@st-andrews.ac.uk.}
\affiliation{SUPA, School of Physics and Astronomy, University of St Andrews, North Haugh, St Andrews, KY16 9SS, United Kingdom}
\author{Peter Wahl}
\email[Correspondence to: ]{wahl@st-andrews.ac.uk.}
\affiliation{SUPA, School of Physics and Astronomy, University of St Andrews, North Haugh, St Andrews, KY16 9SS, United Kingdom}
\affiliation{Physikalisches Institut, Universität Bonn, Nussallee 12, 53115 Bonn, Germany}

\date{\today}

\begin{abstract}
Magnetic van der Waals materials are an important building block to realize spintronic functionalities in heterostructures of two-dimensional (2D) materials. Yet, establishing their magnetic and electronic properties and the interrelationship between the magnetic ground state and electronic structure is often challenging because only a limited number of techniques can probe magnetism and electronic structure on length scales of tens to hundreds of nanometers. Chromium chalcogenides are a class of 2D magnetic materials for which a rich interplay between structure and magnetism has been predicted. Here, we combine angle-resolved photoemission and quasi-particle interference imaging to establish the electronic structure of a monolayer of \ce{CrTe2} on graphite. From a comparison of model calculations with spectroscopic mapping using angle-resolved photoemission spectroscopy and scanning tunnelling microscopy we establish the magnetic ground state and the low energy electronic structure. We demonstrate that the band structure of monolayer \ce{CrTe2} is captured well by density functional theory (DFT) in a DFT+U framework when a Coulomb repulsion of $U=2.5\mathrm{eV}$ is accounted for. 
\end{abstract}


\maketitle

\section{Introduction}
The discovery of magnetic two-dimensional (2D) van der Waals materials \cite{gong_discovery_2017,huang_layer-dependent_2017} has been a milestone in our ability to create heterostructures with spintronic functionalities. Being able to stack materials with strong spin-orbit coupling and magnetic order or monolayers of superconducting and magnetic materials is a prerequisite to creating spin-valve structures and a route to stabilizing unconventional superconducting order parameters \cite{burch_magnetism_2018}. 
The chromium tellurides are a family of 2D magnetic materials which show magnetism at room temperature \cite{sun_room_2020} and for which density functional theory (DFT) predicts a wide range of magnetic ground states dependent on structural details \cite{Wu_CrSe2/Te2_strain, lv_strain-controlled_2015}, including non-collinear magnetic orders \cite{abuawwad_noncollinear_2022} and topological magnetism \cite{abuawwad_crte_2023, abuawwad_electrical_2024}. The family also hosts a range of structurally similar compounds in which ferromagnetic and antiferromagnetic ground states are stabilized with comparatively small changes in the stoichiometry \cite{kushwaha_ferromagnetic_2024}. This range of magnetic ground states makes them an ideal platform to tailor the magnetic order for specific applications. The localized nature of the Cr $d$-orbitals suggests that electronic correlation effects beyond DFT are important \cite{zhu_insight_2023}. Determining their influence on the properties and to assess the validity of the predictions requires detailed knowledge of the electronic structure.

The techniques that are able to detect and characterize magnetism in monolayer materials are limited: magnetization measurements are typically not sufficiently sensitive to pick up a signal from monolayer materials, and neutron scattering requires sample masses on the order of grams. This calls for microscopic techniques that can probe both the magnetism as well as the underpinning electronic structure on sub-micron length scales and in atomically thin materials.
Spin-polarized scanning tunneling microscopy (STM) is one such technique, which can detect magnetism and magnetic order with atomic resolution \cite{wiesendanger_spin_2009,enayat_real-space_2014}. The ability to relate magnetic properties to details of the electronic structure -- crucial to developing a microscopic understanding and theories that can realistically describe the magnetism -- requires in addition the ability to establish the low-energy electronic structure, determined using momentum space techniques. Angle-resolved photoemission spectroscopy (ARPES) is the most commonly used technique for this. Significant progress has been made towards reducing the probing light spot size. Nonetheless, when integrating over multiple flakes or islands, rotational disorder or the presence of multiple domains on small length scales can still make the interpretation challenging.

Here, we demonstrate that the combination of ARPES and quasi-particle interference (QPI) imaging of the 2D van der Waals magnet \ce{CrTe2} enables determination of the detailed low energy electronic structure. This is facilitated by comparison of continuum local density of states calculations for realistic modeling of the QPI and the unfolded spectral function for ARPES with DFT calculations, enabling a detailed account of the electronic structure. We show that the electronic structure exhibits clear fingerprints of the magnetic configuration, which depends sensitively on the in-plane lattice constant as shown by theory \cite{Wu_CrSe2/Te2_strain, lv_strain-controlled_2015, zhu_insight_2023, gao_magnetic_2021, elrashidy_magnetic_2024} and experiment \cite{Miao_tuning_magnetism_2024}. The comparison between calculations and experiment demonstrates that the electronic structure is captured well once a Coulomb repulsion of $2.5\mathrm{eV}$ for the $d$-orbital of \ce{Cr} is included in a DFT+U framework.

\section{Methods}

To grow monolayer films of \ce{CrTe2} by molecular beam epitaxy (MBE), we employ an optimized growth procedure that uses Ge as a sacrificial species to achieve large scale growth of monolayer \ce{CrTe2} \cite{rajan_epitaxial_2024}. Because of close proximity of the growth windows for \ce{CrTe2}, \ce{Cr2Te3} and other intercalated \ce{Cr_xTe_y} compounds, careful control of the growth temperature is crucial \cite{kushwaha_ferromagnetic_2024}. Growth of \ce{CrTe2} is optimized by in-situ cross-checking of the lattice constant using reflection high-energy electron diffraction (RHEED): \ce{CrTe2} has a significantly smaller lattice constant compared to, e.g., \ce{Cr2Te3}. Samples are transported under vacuum to a home-built ultra-high vacuum scanning tunnelling microscope (STM) and measured at $1.6\mathrm{K}$. ARPES data was acquired at the I05 beamline at Diamond Light Source, at temperatures between $40\mathrm{K}$ and $90\mathrm{K}$. We have performed supporting DFT calculations using Quantum Espresso and VASP. For further details, see Supplementary Material.

\section{Results}

\begin{figure}
\centering
\includegraphics[scale=0.5]{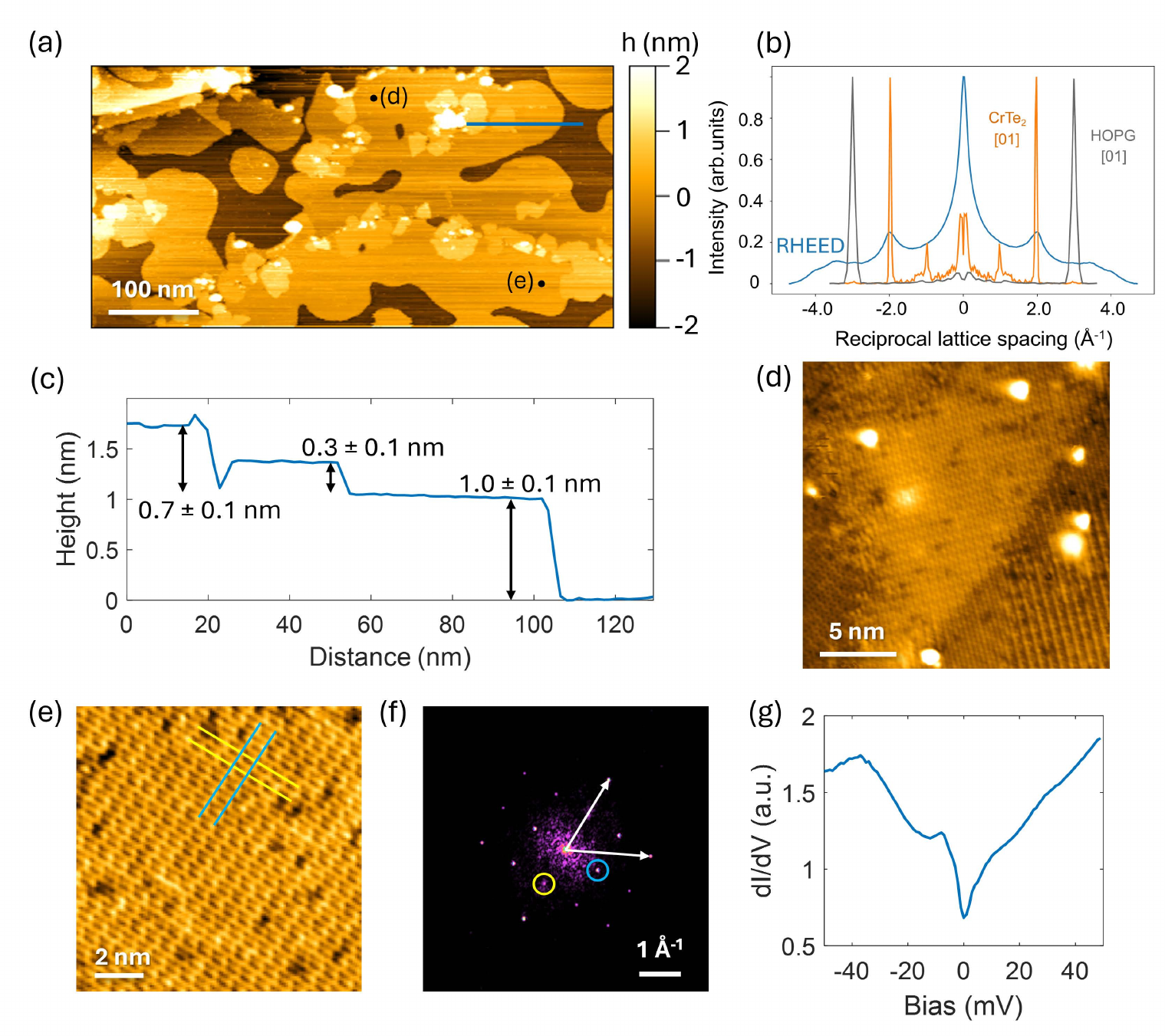}
\caption{\textbf{Growth, structure and morphology of \ce{CrTe2} films.} (a) Large-scale topography of \ce{CrTe2} growth on highly-oriented pyrolytic graphite (tunnelling setpoint $V_\mathrm{s} = 900 \mathrm{mV}$, $I_\mathrm{s} = 60 \mathrm{pA}$), (b) Linecut from RHEED pattern (blue) with linecut through Fourier transformations (FTs) of topographic images of \ce{CrTe2} (orange) and HOPG (grey) from STM. (c) Linecut along line shown in topography (a) ($V_\mathrm{s} = 900 \mathrm{mV}$, $I_\mathrm{s} = 60 \mathrm{pA}$). (d) Topography on a \ce{CrTe2} island showing domains of zig-sag/stripe order ($V_\mathrm{s} = 200 \mathrm{mV}$, $I_\mathrm{s} = 200 \mathrm{pA}$). (e) Atomic-resolution image of a single domain of the zig-zag order ($V_\mathrm{s} = 250 \mathrm{mV}$, $I_\mathrm{s} = 250 \mathrm{pA}$) and (f) the corresponding Fourier transformation, with atomic peak positions marked by the white arrows. The stripe and zig-zag modulations are shown in yellow and blue. (g) Differential conductance spectrum $g(V)$ of monolayer \ce{CrTe2} ($V_\mathrm{s} = 50 \mathrm{mV}$, $I_\mathrm{s} = 100 \mathrm{pA}$, $V_\mathrm{m}=0.8 \mathrm{mV}$, average of 256 spectra from a spectroscopic map).}
\label{fig:STM_topos}
\end{figure}

A typical topographic image following growth of \ce{CrTe2} on highly-oriented pyrolytic graphite (HOPG) is shown in Fig.~\ref{fig:STM_topos}(a), with monolayer-high islands and only very low coverage of the second layer. The regions between the islands are the bare graphite substrate. Visibility of both graphite and \ce{CrTe2} in the same image allows calibration of the lattice constants between the two, giving a lattice constant of $3.55 \pm 0.15 \mathrm{\AA}$ for the \ce{CrTe2} islands, slightly smaller but consistent with the lattice constant found by RHEED for the \ce{CrTe2} layer, which is $3.67 \pm 0.02 \mathrm{\AA}$ (Fig.~\ref{fig:STM_topos}(b)). 

The line cut in Fig.~\ref{fig:STM_topos}(c) shows that the height of the islands is consistent with monolayer \ce{CrTe2}, which for the first layer has an apparent height of about $1\mathrm{nm}$. Although larger than the $c$ parameter of the bulk \ce{CrTe2}, which is 6.10\AA{ }\cite{freitas2015ferromagnetism}, this height is consistent with previous STM and scanning transmission electron microscopy (STEM) measurements \cite{xian2022spin, Miao_tuning_magnetism_2024}. We compare this to the apparent step height calculated using DFT which gives a value for the height of a monolayer of $1$nm (see Supplementary Material, Section \ref{supp:subsection:DFT} for details), in good agreement with the experimental value. 

Images acquired on individual islands reveal a zig-zag order superimposed on the atomic resolution (Fig.~\ref{fig:STM_topos}(d) and (e)), consistent with previous studies and suggesting a zig-zag antiferromagnetic (zz-AFM) order \cite{xian2022spin, Miao_tuning_magnetism_2024}. The magnetic ground state of monolayer \ce{CrTe2} is predicted to depend sensitively on the lattice constant \cite{Wu_CrSe2/Te2_strain, lv_strain-controlled_2015, zhu_insight_2023, gao_magnetic_2021, elrashidy_magnetic_2024}. For the value measured here, the ground state is expected to be antiferromagnetic. The zig-zag order was previously identified as an antiferromagnetic order from spin-polarized STM \cite{xian2022spin, Miao_tuning_magnetism_2024}. It has twice the periodicity of \ce{CrTe2} and breaks the six-fold rotational symmetry (see Fig.~\ref{fig:STM_topos}(f)). We observe multiple domains of the order within patches of $\sim 20\mathrm{nm}$ (as shown in Fig.~\ref{fig:STM_topos}(d)). An additional stripe order, with the same periodicity as the zig-zag order but in the perpendicular direction, is also present. In agreement with previous studies \cite{xian2022spin, Miao_tuning_magnetism_2024}, we attribute this order to a small distortion of the lattice (see Suppl. Fig.~\ref{supp:fig:Topo_linecuts}). Spatially-averaged tunneling spectra recorded in the vicinity of the Fermi energy show that the islands are metallic \cite{kushwaha_ferromagnetic_2024}, with a dip in the spectra close to zero bias, and a peak-like feature at $-40\mathrm{mV}$ (Fig.~\ref{fig:STM_topos}(g)).

\begin{figure}
\centering
\includegraphics[scale=0.45]{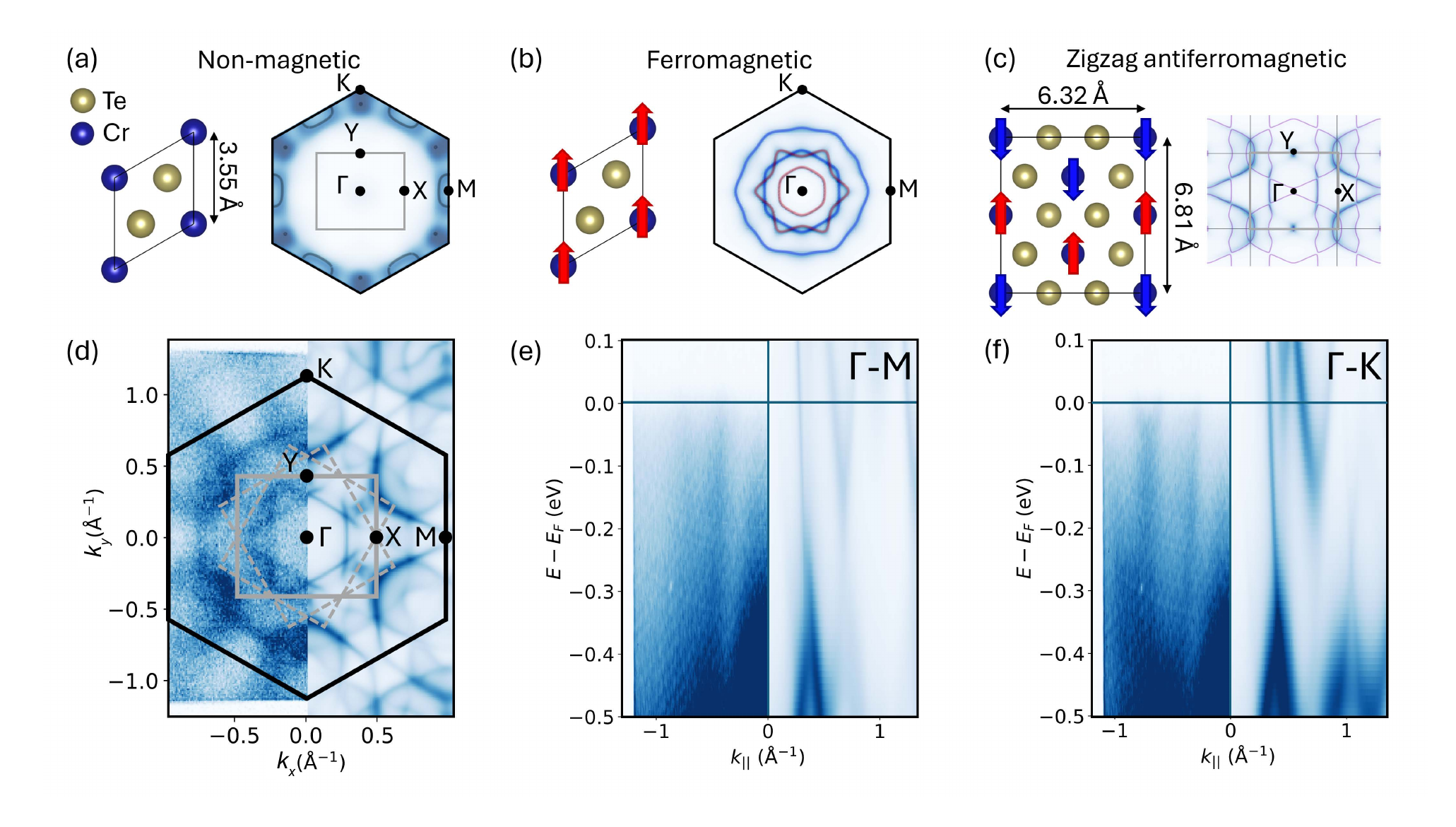}
\caption{\textbf{Electronic structure from ARPES and DFT calculations.} (a)-(c) Left: unit cell, right: Fermi surface overlaid on calculated spectral function for (a) non-magnetic \ce{CrTe2}, (b) ferromagnetic and (c) zig-zag antiferromagnetic (zz-AFM) order, obtained from DFT calculations for a free-standing monolayer. For the ferromagnetic case, red and blue indicate spin up and down bands. The grey rectangle in (a) indicates the zz-AFM Brillouin zone (all calculations with $U=2.5\mathrm{eV}$). (d) Left: ARPES constant energy contour acquired at the Fermi energy, averaged between $\pm25 \mathrm{meV}$. Right: unfolded spectral function calculated for the zz-AFM order with an energy broadening of $100 \mathrm{meV}$ and a shift of the chemical potential by $50 \mathrm{meV}$ in the calculation. The spectral function is six-fold symmetrized to account for rotational domains of the zz-AFM order. The hexagonal Brillouin zone is shown in black; the zz-AFM Brillouin zone and its rotated copies are shown in grey. (e) and (f) Cuts along the ${\Gamma}$-M and $\Gamma$-K directions from the ARPES data (left half) and spectral function (right half) for the calculation for the zz-AFM order (again symmetrized and including the same shift as in (d)).}
\label{fig:FS_ARPES}
\end{figure}

To establish a detailed understanding of the electronic structure, we have performed DFT+$U$ calculations for non-magnetic, ferromagnetic and zig-zag antiferromagnetic configurations of a free-standing monolayer of \ce{CrTe2}. For all the magnetic configurations investigated here (non-magnetic, ferromagnetic and antiferromagnetic), we find Fermi surfaces with multiple bands, predominantly of Te character, crossing the Fermi energy (Fig.~\ref{fig:FS_ARPES}(a)-(c), all with $U=2.5\mathrm{eV}$, see also Suppl. Fig.~\ref{supp:fig:DFT_band_structure}(a), (b) and (f)). The electronic structure for the antiferromagnetic configuration exhibits significant differences from the ferromagnetic and non-magnetic ones, not least because of the different symmetry of the Brillouin zone (shown in Fig.~\ref{fig:FS_ARPES}(a)).

The Fermi surface as measured by angle-resolved photoemission of monolayer \ce{CrTe2} on natural graphite (Fig.~\ref{fig:FS_ARPES}(d)) confirms the metallic character. It has apparent six-fold symmetry and consists of a sharper inner hexagon surrounded by triangular features and a larger broader hexagon, both of which are centred around the $\Gamma$-point. Around the $K$-points in the hexagonal Brillouin zone there are broader pockets with high intensity. These last two features can also be seen in measurements taken with a higher photon energy (Suppl. Fig.~\ref{supp:fig:ARPES_90eV}), as well as some intensity at the corners of the inner hexagon. 

Comparison with the calculations shows poor agreement with the non-magnetic case as there are no features in the centre of the Brillouin zone for such a calculation: the Fermi surface in Fig.~\ref{fig:FS_ARPES}(a) consists of small pockets around the $M$- and $K$-points and the intensity of the calculated spectral function is highest around the zone edges . In the ferromagnetic case, all bands are centred around the $\Gamma$-point (Fig.~\ref{fig:FS_ARPES}(b)), again inconsistent with the ARPES measurement, which also shows pockets enclosing the $K$-points. The zz-AFM configuration, shown in Fig.~\ref{fig:FS_ARPES}(c), enlarges the \ce{CrTe2} unit cell and therefore reduces the size of the Brillouin zone. For an improved comparison with ARPES, we have unfolded the band structure obtained from DFT to calculate an unfolded spectral function (see Supplementary Material, Section \ref{supp:subsection:unfolded} for details), however neglecting, e.g., self-energy effects beyond an energy-independent broadening. The most notable effect of the unfolding is that the bands closest to the $\Gamma$-point retain very little spectral weight: their dominant spectral weight is in higher order Brillouin zones. 

The antiferromagnetic configuration also changes the symmetry of the unit cell, from six- to two-fold. The Fermi surface measured in ARPES shows clear hexagonal symmetry. However, because the $\sim5\mathrm{\mu m}$ spot size of the incident beam is much larger than both the magnetic domain size and the island size (as shown in Fig.~\ref{fig:STM_topos}(d) and (a))), photoelectrons will stem from multiple islands on which the magnetic order forms independently, thus averaging over different domains of the zigzag antiferromagnetic order. 
To compare the DFT calculations and the ARPES data, we therefore account for the magnetic domains of the zz-AFM order by symmetrizing the calculation, adding copies of the unfolded spectral function rotated by $60$\degree{ }and $120$\degree.
Fig.~\ref{fig:FS_ARPES}(d) shows the spectral function obtained after such a symmetrization (with $U = 2.5 \mathrm{eV}$) in comparison with the Fermi surface obtained from ARPES. The calculated spectral function shows very good agreement: the inner hexagon and triangles as well as the higher-intensity arcs near the $K$-points are consistently reproduced in the calculations. 

To assess the impact of electron correlations, we compared the data with calculations performed with a range of $U$ values between $0$ and $6 \mathrm{eV}$. While the precise value of $U$ has relatively little influence on the band structure, there are notable differences between $U=0\mathrm{eV}$ and non-zero values of $U$. For example, an additional Van Hove singularity, which is not observed experimentally, is found below the Fermi energy at the $\Gamma$ point for $U=0\mathrm{eV}$ (see Suppl. Fig.~\ref{supp:fig:DFT_band_structure}). The inclusion of the $U$-term also causes the Cr-derived bands to be pushed away from the Fermi energy.
The results of the spectral function calculations demonstrate that the inclusion of a non-zero $U$ value improves the agreement with experiment (see Suppl. Fig.~\ref{supp:fig:ARPES_U_comparison} for comparisons with various $U$ values). While key features of the ARPES data are recognizable for all values of $U$, their sizes and locations vary with the precise value. We find a $U$ value of around $2$ to $3 \mathrm{eV}$ produces results most consistent with the experimental data and use $U=2.5\mathrm{eV}$ for all comparisons and calculations. We also allow for an energy shift in the calculations for comparison with the data to account for charge transfer between the substrate and the \ce{CrTe2} monolayer, which differs slightly for each $U$ value used in the calculation. In the case of $U=2.5\mathrm{eV}$, the results of the calculations are shifted upwards in energy by $50\mathrm{meV}$, resulting in a good agreement with the ARPES measurements. In contrast, the non-magnetic and ferromagnetic spectral functions with $U = 2.5\mathrm{eV}$ (Fig.~\ref{fig:FS_ARPES}(a) and (b)) are very different from the experimental Fermi surface. Notably, while a $U$-term is required to account for correlations on the Cr-site, comparison of the DFT band structure with ARPES shows that no further renormalization of the band structure is required (compare Suppl. Fig.~\ref{supp:fig:ARPES_renorm}). This is likely a consequence of the states crossing the Fermi energy being dominantly Te-derived states.

This agreement of the calculated spectral function with ARPES for a zz-AFM configuration is supported by comparing line cuts through the symmetrized spectral function with the measured photoemission dispersions (Fig.~\ref{fig:FS_ARPES}(e) and (f)). Hole-like bands from the inner region of the Fermi surface and the broader outer hexagon are visible in both experiment and calculation. The features around the $K$-points in Fig.~\ref{fig:FS_ARPES}(d) are much sharper in the spectral function calculation, but the band minimum near the Fermi energy in the $\Gamma$-K direction (Fig.~\ref{fig:FS_ARPES}(f)) is broadly consistent with a similar feature in the measured ARPES data.

\begin{figure}
\centering
\includegraphics[scale=0.65]{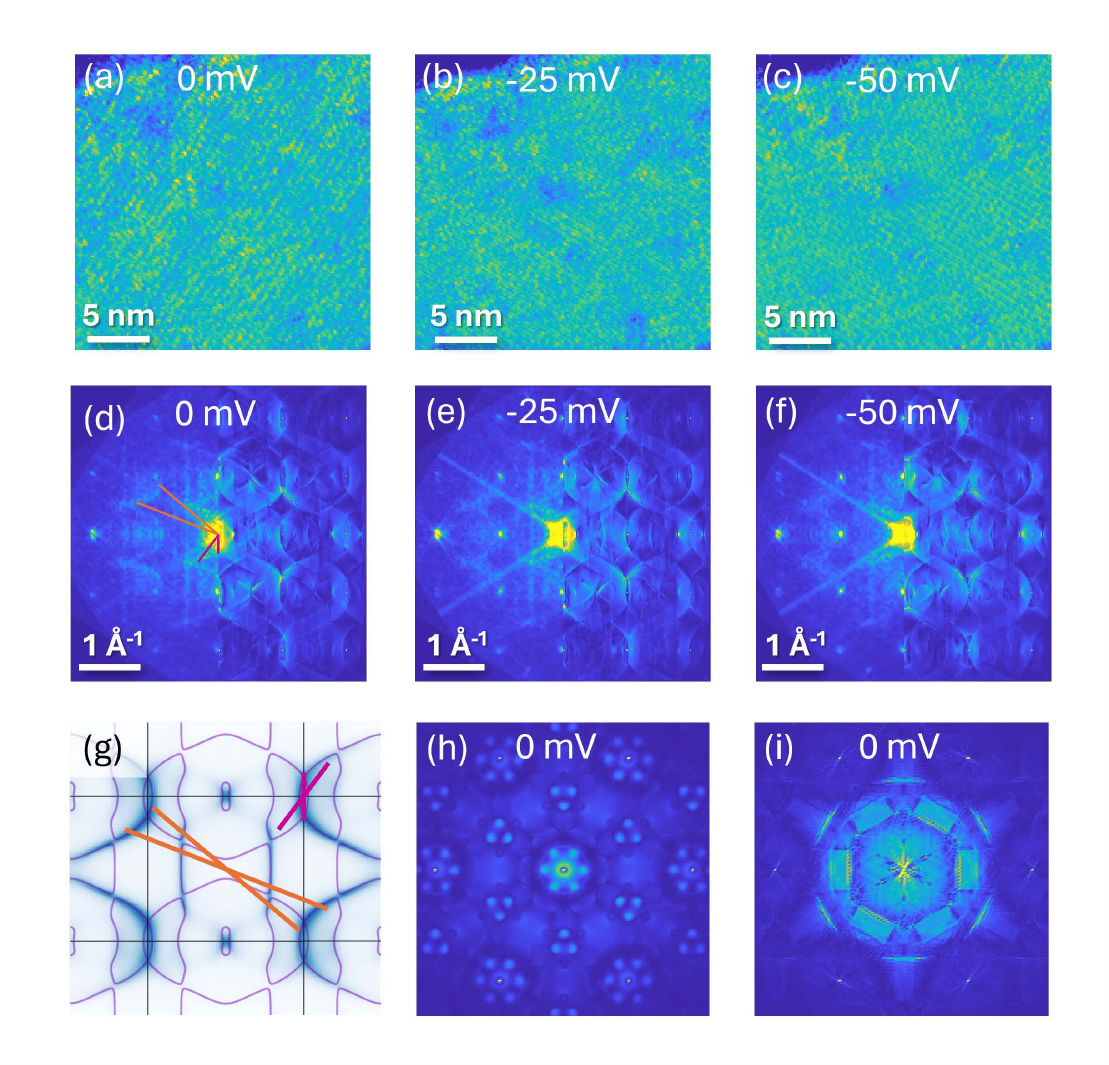}
\caption{\textbf{Quasi-particle interference imaging of \ce{CrTe2}.} (a)-(c) Real space differential conductance images of spectroscopic map layers at (a) $0 \mathrm{mV}$, (b) $-25 \mathrm{mV}$ and (c) $-50 \mathrm{mV}$. (d)-(f) Left: Fourier transform of (a)-(c) symmetrized using two mirror planes (horizontal and vertical). Right: Fourier transforms of continuum local density of states (CLDOS) calculations at the same energies for a zz-AFM unit cell (with a negative scattering potential, $U = 2.5$eV and including a $50\mathrm{meV}$ energy shift). Map setpoint: $V_\mathrm{s} = 100 \mathrm{mV}$, $I_\mathrm{s} = 200 \mathrm{pA}$, lock-in amplitude $4 \mathrm{mV}$. The diagonal lines are artefacts of the Fourier transform. (g) Scattering vectors from (d) on a constant energy contour and unfolded spectral function calculation, both at the Fermi energy (with $U = 2.5$eV and including $50\mathrm{meV}$ energy shift). (h) and (i) Fourier transforms of CLDOS calculations at $0 \mathrm{meV}$ with a positive scattering potential and $U=2.5\mathrm{eV}$ for a (h) non-magnetic unit cell and (i) ferromagnetic unit cell.}
\label{fig:QPI}
\end{figure}

We use quasi-particle interference (QPI) imaging to further refine the electronic structure. QPI allows selecting mono-domain areas in the sample, such as the one shown in Fig.~\ref{fig:STM_topos}(e), and therefore facilitates disentangling the electronic structure within an individual domain. The QPI obtained from a spectroscopic map is shown in Fig.~\ref{fig:QPI}. In Fig.~\ref{fig:QPI}(a)-(c), we show the real-space maps of the QPI, and in the left halves of Fig.~\ref{fig:QPI}(d)-(f) their Fourier transforms (after symmetrization using a horizontal and a vertical mirror plane). The  Fourier transform reveals four sharp peaks suggesting a rectangular symmetry of the Brillouin zone and distinct patterns from the quasi-particle interference itself.  The QPI patterns disperse strongly in energy, confirming their electronic origin. The right halves of Fig.~\ref{fig:QPI}(d)-(f) show Fourier transforms of QPI patterns calculated using the continuum Green's function approach (see Supplementary Material, Section \ref{supp:subsection:CLDOS} for details) based on the electronic structure with zz-AFM order using the Hubbard $U$ term of $2.5\mathrm{eV}$ and an energy shift of $50\mathrm{meV}$ as obtained from the comparison with the ARPES measurements. The choice of $U$ value from this comparison is supported by the good agreement here between experimental and calculated QPI, when compared with other $U$ values (shown in Suppl. Fig.~\ref{supp:fig:QPI_U_comparison}). The arcs in the Fourier transform are consistent with the shape of the zz-AFM Fermi surface, as indicated by the scattering vectors in Fig.~\ref{fig:QPI}(d) and (g). These regions of the Fermi surface also have the highest intensity in the calculated spectral function and correspond to the pockets around the $K$-points of the hexagonal Brillouin zone in the ARPES measurements (Fig.~\ref{fig:FS_ARPES}(d)), enabling the direct relation of QPI and ARPES. We note that details of the scattering patterns depend also on the precise impurity potential (compare Suppl. Fig.~\ref{supp:fig:Defect_comparison} for calculations with different defect positions and Suppl. Fig.~\ref{supp:fig:Potential_comparison} with positive potential).

Considering the quasi-particle interference patterns expected for the three ground states introduced in Fig.~\ref{fig:FS_ARPES}, the symmetry of the QPI patterns for the non-magnetic and ferromagnetic states (Fig.~\ref{fig:QPI}(h) and (i)) is notably different compared to the zz-AFM state due to the symmetry of the magnetic order. This shows that not only does comparing the simulated QPI enable us to determine the low energy electronic structure, but that it also provides information about the unit cell and hence magnetic order.

\begin{figure}
\centering
\includegraphics[scale=0.55]{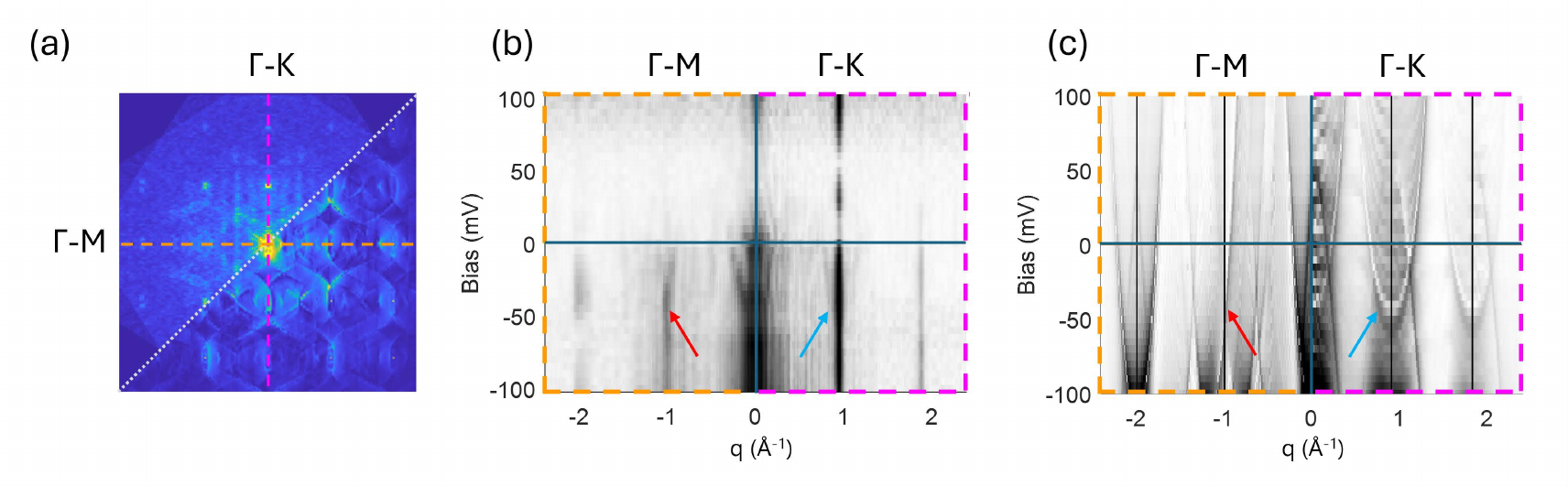}
\caption{\textbf{Comparison of QPI and DFT.} (a) QPI layer $\tilde{g}(\mathbf{q},eV)$ (upper left half) and continuum local density of states (CLDOS) calculation (lower right half) based on the tight-binding model adjusted to the ARPES data. Pink and orange dashed lines show the high-symmetry directions of the Brillouin zone. (b) and (c) Cut through the measured STM data (b) and the CLDOS calculation using the optimized model for the electronic structure (see main text) (c) along the same path in reciprocal space, including an energy shift of $50 \mathrm{meV}$. Corresponding dispersions are marked with red and blue arrows. Map setpoint: $V_\mathrm{s} = 100 \mathrm{mV}$, $I_\mathrm{s} = 100 \mathrm{pA}$, lock-in amplitude $4 \mathrm{mV}$. Cuts shown in (b) are averages of five slices around the centre of the image.} 
\label{fig:Bands}
\end{figure}

A detailed comparison of the experimental and calculated QPI for the zz-AFM order is shown in Fig.~\ref{fig:Bands}. We consider cuts along the high symmetry directions, indicated in Fig.~\ref{fig:Bands}(a), from both the experiment, Fig.~\ref{fig:Bands}(b), and calculations, Fig.~\ref{fig:Bands}(c). Key features from the cuts can be seen in the calculations, as marked by arrows, and correspond to the arc features moving towards the Bragg peaks as a function of energy. The minimum of these features in the $\Gamma$-M direction in Fig.~\ref{fig:Bands}(c) corresponds to the minimum of the electron-like band in ARPES at about $-100\mathrm{meV}$ (see Fig.~\ref{fig:FS_ARPES}(f)) and where the pockets around the $K$-points in the ARPES data disappear and become incoherent. The calculations clearly show more and sharper features than observed experimentally. This is likely a consequence of neglecting electronic correlation effects in the calculations, which in the experiment result in an energy-dependent finite lifetime and decoherence of the electronic states. We note that our calculations also account for scattering only from the Cr site, whereas in reality, there will also be defects on the Te sites resulting in different intensities of the scattering patterns (see Suppl. Fig. ~\ref{supp:fig:Defect_comparison} for comparison with scattering from a defect on the Te sites). Overall, however, key features are reproduced, with no band renormalization required, consistent with the analysis of the ARPES data shown above.

\section{Discussion}
The comparison between band structure calculations, QPI modelling, ARPES and STM allows us to identify the electronic structure of the monolayer as that of zig-zag antiferromagnetically ordered \ce{CrTe2}, and hence confirms the ground state seen also in spin-polarized STM studies of monolayer \ce{CrTe2} \cite{xian2022spin, Miao_tuning_magnetism_2024}. It confirms the importance of the $U$ term of $U=2.5\mathrm{eV}$, lower than values derived from linear response methods \cite{liu_crte2_2022, Wu_CrSe2/Te2_strain}, required to accurately model the electronic structure. The need for including this $U$ term suggests that electron correlations cannot be neglected and that they also play an important role for the magnetic properties. Nonetheless, the relatively small $U$ value which we find gives the optimum comparison to the experimental data points to the strong Cr-ligand hybridisation, in line with other studies of Te-based 2D magnets \cite{frisk_magnetic_2018,watson_direct_2020}. The magnetic phase diagram with small structural distortions has already been found to be very rich from DFT calculations performed without inclusion of a $U$ term \cite{abuawwad_crte_2023}.  
Our work establishes \ce{CrTe2} as a promising platform to study the interplay between magnetism in two dimensions. Inclusion of spin-orbit coupling, which is sizeable for the Te p-bands, does not have a significant impact on the electronic structure close to the Fermi energy here, so that it can be neglected in the comparison between theory and experiment for the bands close to the Fermi energy (compare suppl. fig.~\ref{supp:fig:soc-comparison}). 
Our results further establish QPI as a tool to study the electronic structure of monolayer transition metal dichalcogenides, and show that combination with continuum QPI calculations and comparison with ARPES allows extraction of detailed information about the low energy electronic band structure in these systems. QPI is particularly useful here as a complementary tool to angle-resolved photoemission, because of the possibility of selecting individual magnetic domains.
The electronic structure we obtain is remarkably consistent with the prediction by DFT for a free-standing monolayer. While a non-negligible Coulomb repulsion $U$ must be included, it is significantly smaller than required to describe the electronic states, e.g., in chromium oxides\cite{le_magnetic_2018,sunko_probing_2020}. This is likely because the electronic states close to the Fermi energy have dominant Te character, whereas the Cr bands are pushed away from the Fermi energy by the Coulomb repulsion.

\section{Conclusion}
Through the combined application of angle-resolved photoemission and quasi-particle interference imaging, we can resolve the low energy electronic structure of 2D magnetic systems. Detailed comparison with calculations shows that correlation effects are important in this system, reflected in a non-negligible $U$ term required to obtain a good agreement between calculations and experiment. This suggests that also modelling of magnetism will require correlation effects to be taken into account, given the sensitivity of the magnetic ground state to structural details. The detailed understanding of the electronic structure opens the door to a designer approach to 2D magnetism. 

\nocite{rajan_epitaxial_2024,kushwaha_ferromagnetic_2024,Giannozzi_QE_2009, Giannozzi_QE_2017,PBE,grimme_dftd3_2010,Tersoff-Hamann,kresse_ab_1993,kresse_ab_1994,kresse_norm-conserving_1994,kresse_efficiency_1996,kresse_efficient_1996,kresse_ultrasoft_1999,kushwaha_ferromagnetic_2024,benedicic_interplay_2022,rhodes_nature_2023,marques_tomographic_2021,naritsuka_compass-like_2023,marques_spin-orbit_2024,choubey_visualization_2014,kreisel_interpretation_2015,kreisel_quasi-particle_2021}

\label{Bibliography}
\bibliographystyle{unsrtnat}
\bibliography{crte2}
\end{document}



\title{Supplementary Material for 'Electronic structure of monolayer-\ce{CrTe2}: an antiferromagnetic 2D van der Waals material'}


\author{Olivia Armitage}
\affiliation{SUPA, School of Physics and Astronomy, University of St Andrews, North Haugh, St Andrews, KY16 9SS, United Kingdom}
\author{Naina Kushwaha}
\affiliation{SUPA, School of Physics and Astronomy, University of St Andrews, North Haugh, St Andrews, KY16 9SS, United Kingdom}
\author{Akhil Rajan}
\affiliation{SUPA, School of Physics and Astronomy, University of St Andrews, North Haugh, St Andrews, KY16 9SS, United Kingdom}
\author{Luke C. Rhodes}
\author{Sebastian Buchberger}
\author{Bruno Kenichi Saika}
\author{Shu Mo}
\affiliation{SUPA, School of Physics and Astronomy, University of St Andrews, North Haugh, St Andrews, KY16 9SS, United Kingdom}
\author{Matthew D. Watson}
\affiliation{Diamond Light Source Ltd., Harwell Science and Innovation Campus, Didcot OX11 0DE, UK}
\author{Phil D. C. King}
\email[Correspondence to: ]{pdk6@st-andrews.ac.uk.}
\affiliation{SUPA, School of Physics and Astronomy, University of St Andrews, North Haugh, St Andrews, KY16 9SS, United Kingdom}
\author{Peter Wahl}
\email[Correspondence to: ]{wahl@st-andrews.ac.uk.}
\affiliation{SUPA, School of Physics and Astronomy, University of St Andrews, North Haugh, St Andrews, KY16 9SS, United Kingdom}
\affiliation{Physikalisches Institut, Universität Bonn, Nussallee 12, 53115 Bonn, Germany}

\date{\today}


\maketitle

\section{Methods}
\subsection{Thin film growth}
Monolayer \ce{CrTe2} was synthesized on highly oriented pyrolytic graphite (HOPG) and natural graphite substrates using a DCA R450 molecular beam epitaxy (MBE) system, with a base pressure of approximately $\sim\!1\times10^{-10}$~mbar. Chromium (Cr) and tellurium (Te) were evaporated from effusion cells at temperatures of 1025$^\circ$C and 425$^\circ$C, respectively, achieving a Cr-Te flux ratio of approximately 1:200. Before the film growth, the substrates were outgassed in the growth chamber at 800$^\circ$C for approximately 25 minutes. The monolayer CrTe$_2$ was grown with a substrate temperature of 400$^\circ$C for one hour and subsequently cooled in the presence of Te. To enhance nucleation, we  co-evaporated small quantities of Ge from an electron beam evaporator during the growth. The incident flux of Ge ions at the substrate surface enhances the nucleation of the epilayer, promoting higher growth rates and improving uniformity of the grown layer \cite{rajan_epitaxial_2024,kushwaha_ferromagnetic_2024}. In situ reflection high-energy electron diffraction (RHEED) was used to monitor surface quality and phase formation throughout the growth process.

\subsection{DFT calculations}
\label{subsection:DFT}
We performed structural relaxation of monolayer \ce{CrTe2} and a graphene layer to estimate the theoretical step height using Quantum Espresso \cite{Giannozzi_QE_2009, Giannozzi_QE_2017}, using the
Perdew–Burke–Ernzerhof (PBE) exchange-correlation functional \cite{PBE}, projector augmented
wave (PAW) pseudopotentials, and an energy cutoff of $40 \mathrm{Ry} = 544 \mathrm{eV}$. The calculation includes a DFT-D3 correction to account for van der Waals interactions \cite{grimme_dftd3_2010}, but does not include magnetism or a Hubbard $U$, as these are not expected to significantly affect the structure. During the relaxation, the unit cell parameters and atomic positions are allowed to vary until the force on each atom is less than $1 \times 10^{-3}$ Rydberg
atomic units $= 0.01 \mathrm{eV}/\mathrm{\AA}$. For the relaxation, we used a $6 \times 6 \times 1$ $k$ grid and a $30\mathrm{\AA}$ vacuum layer to prevent interactions between adjacent layers. 

The relaxed structure has a monolayer thickness of $\sim3\mathrm{\AA}$ and a van der Waals gap of $\sim4\mathrm{\AA}$. We also estimate the electronic contribution to the island height measured in STM, which is due to the different densities of states of the \ce{CrTe2} and the graphite substrate. To this end, we performed a non-self-consistent field calculation on the relaxed structure with an energy cutoff of $40 \mathrm{Ry}$ and a $k$ grid of $16 \times 16 \times 1$. The density of states integrated over energy in real space is then simulated using the Tersoff-Hamann approximation \cite{Tersoff-Hamann}. This quantity is proportional to the tunnelling current. We then calculate its average over the plane parallel to the layers and plot this as a function of distance from the layer surface. 
By comparing the distances at which a threshold value is reached between a \ce{CrTe2} monolayer and graphene layer, we estimate that the measured step height increases by $\sim3$\AA. Combining these contributions gives a calculated height of $1\mathrm{nm}$ for monolayer \ce{CrTe2}.

Further band structure calculations for different values of the Coulomb repulsion $U$ and for projection onto tight-binding models for the QPI calculations have been done using VASP with the same structure of a free-standing \ce{CrTe2} layer as in the calculations with Quantum Espresso. The calculations have been performed using the PBE functional \cite{kresse_ab_1993,kresse_ab_1994,kresse_norm-conserving_1994,kresse_efficiency_1996,kresse_efficient_1996,kresse_ultrasoft_1999} and with a plane wave energy cut off of $600\mathrm{eV}$, a $\mathbf{k}$-point grid of $8\times 8\times 1$, and using $U$ as indicated in the main text and $J=0.3U$. For the continuum Green's function calculation, we project the band structure and wave functions onto localized orbitals using a modified version of Wannier90 to obtain a tight-binding model as well as the localized wave functions to describe the vacuum overlap.

\subsection{ARPES}
ARPES measurements were performed at the nano-ARPES branch line of the I05 beamline at Diamond Light Source, UK. The light was focussed to a spot of $\approx\!5$~$\mu$m using a capillary mirror. The data presented in Fig. \ref{main:fig:FS_ARPES}(d)–(f) in the main text were acquired at a photon energy of $62\mathrm{eV}$ with both linearly horizontal (LH) and vertical (LV) polarizations and are plotted here as the sum of the two spectra measured with each polarization (LH+LV). The measurements were performed at temperatures of $40\mathrm{K}$ and $90\mathrm{K}$, well below the magnetic transition temperature of monolayer \ce{CrTe2} \cite{kushwaha_ferromagnetic_2024}. The total energy resolution, including contributions from both the analyzer and the beamline, was approximately $40\mathrm{meV}$. For these measurements, the monolayer \ce{CrTe2} was grown on a natural graphite substrate and transferred from the MBE to the ARPES system via a vacuum suitcase to prevent any exposure to air. 

\subsection{STM}
STM measurements were performed in a home-built STM that operates in ultra-high vacuum (UHV) at a temperature of $1.6\mathrm K$. Samples are transferred in a vacuum suitcase from the MBE system to the STM, and are not usually exposed to a pressure of more than $5\cdot 10^{-9}\mathrm{mbar}$. The bias voltage $V$ is applied to the sample. Tunneling spectra are acquired in open feedback loop condition after stabilizing the tip height at set point conditions ($V_\mathrm s$, $I_\mathrm s$), recording the differential conductance $g(V)$ using a lock-in amplifier with a lock-in modulation $V_\mathrm{m}$ added to the sample bias $V$. 

\subsection{Continuum Green's function calculations}
\label{subsection:CLDOS}
We perform continuum Green's function calculations using the St Andrews calcqpi code\cite{benedicic_interplay_2022,rhodes_nature_2023,naritsuka_compass-like_2023,marques_spin-orbit_2024}, which uses the formalism developed by Hirschfeld \textit{et al.}\cite{choubey_visualization_2014,kreisel_interpretation_2015,kreisel_quasi-particle_2021} to model the quasi-particle interference from a defect. The QPI is calculated on a real-space grid of $64\times 64$ unit cells using $128\times128$ $k$-points. For all QPI calculations, we have used a broadening $\eta=5\mathrm{meV}$. Due to the two-dimensional nature of the sample, interlayer hopping, as would be relevant in bulk samples\cite{marques_tomographic_2021,rhodes_nature_2023}, does not result in additional broadening here.

\subsection{Unfolded spectral function}
\label{subsection:unfolded}
To obtain the unfolded spectral function from a tight-binding model, following ref.~\onlinecite{rhodes_probing_2024}, we calculate
\begin{equation}
A(\mathbf{k},\omega)=-\frac{1}{\pi}\sum_{\mu,\nu}e^{i\mathbf{k}(\mathbf{r}_\mu-\mathbf{r}_\nu)}\cdot\mathrm{Im} G_0^{\mu\nu}(\mathbf{k},\omega),
\end{equation}
where $\mu$, $\nu$ are orbital indices, and $\mathbf{r}_\mu$ are the positions of orbital $\mu$. For the unfolded spectra function, calculations are done on a $\mathbf{k}$-grid of $512\times 512$ points, for the ferromagnetic and non-magnetic calculation distributed across the Brillouin zone, for the antiferromagnetic calculation, $128\times 128$ points are considered on a uniform grid in the Brillouin zone.

\section{Additional data and calculations}
\subsection{STM data}
\begin{figure}
\centering
\includegraphics[scale=0.6]{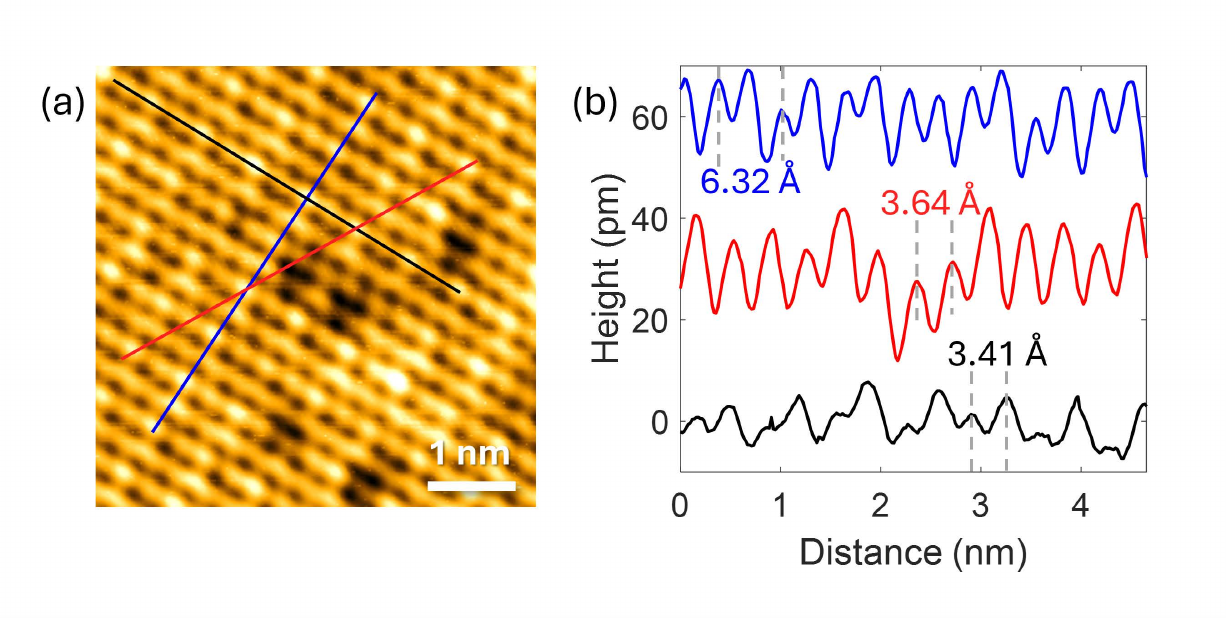}
\caption{Structural distortion of the \ce{CrTe2} lattice. (a) Topographic image, with $V_{\mathrm{S}}=-100 \mathrm{mV}$ and $I_{\mathrm{S}}=100 \mathrm{pA}$, showing the stripe and zig-zag periodicities. (b) Line cuts along the three lines shown in (a) (all left to right), with the average periodicity marked, showing different Te-Te distances in different directions.}
\label{fig:Topo_linecuts}
\end{figure}

In fig.~\ref{fig:Topo_linecuts}, a high-resolution STM image of monolayer \ce{CrTe2} and line cuts show the distortion of the lattice. The difference in periodicities between the red and black lines illustrates the structural distortion. The doubling of the \ce{CrTe2} unit cell is evident in the periodicity of the the measured height in the blue and black lines. 

\subsection{Band structure calculations for different values of $U$}
\begin{figure}
\centering
\includegraphics[width=\textwidth]{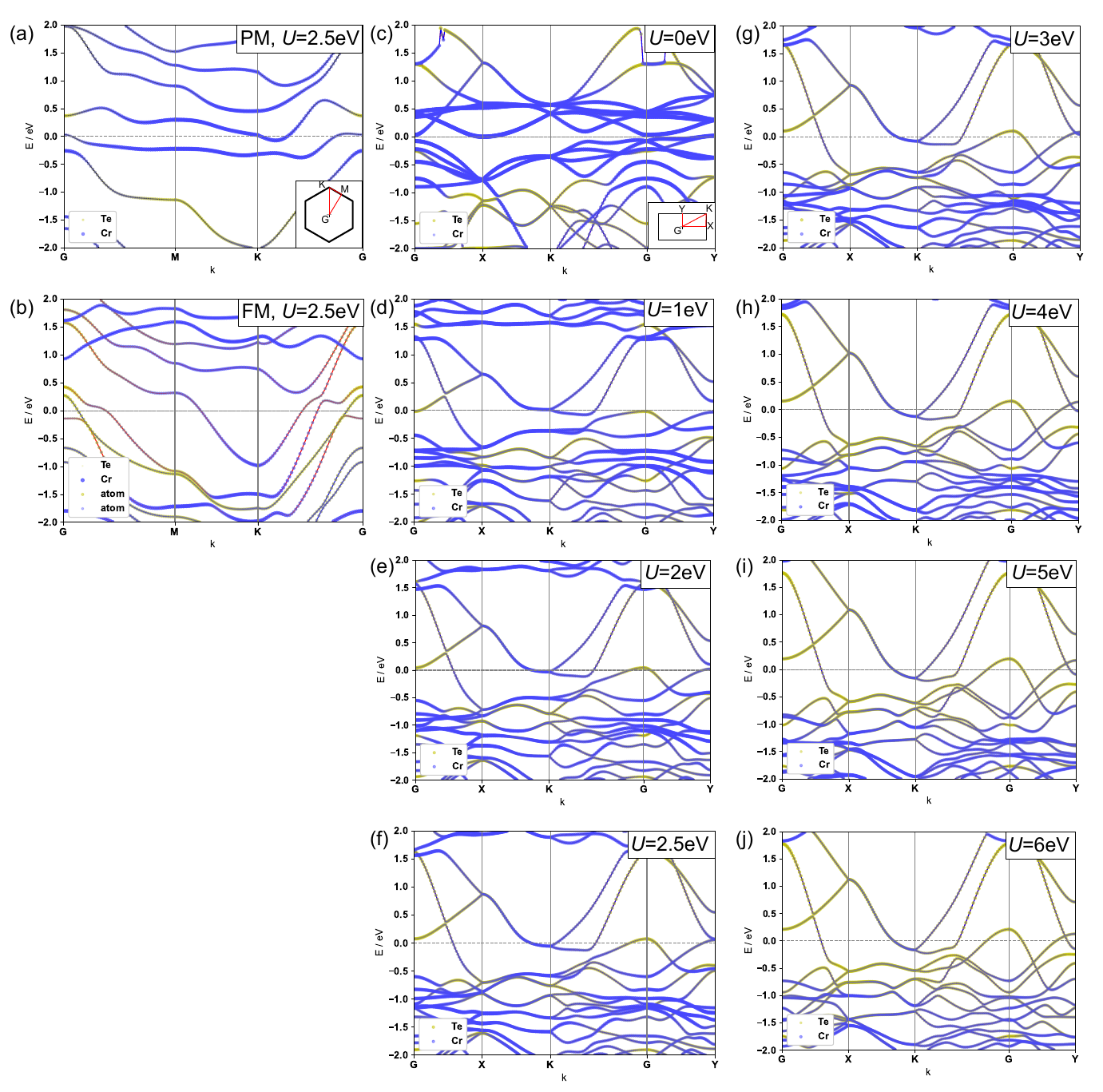}
\caption{Band structure obtained from DFT calculations for a non-magnetic calculation (a) and for a ferromagnetic calculation (b), both with $U=2.5\mathrm{eV}$. Panels (c-j) show calculations for the zig-zag AFM order for $U=0, 1, 2, 2.5, 3, 4, 5, 6\mathrm{eV}$, respectively.}
\label{fig:DFT_band_structure}
\label{fig:Band_structure_U}
\end{figure}
Fig. \ref{fig:Band_structure_U}(a) and (b) show the band structures for the non-magnetic and ferromagnetic hexagonal unit cells, with $U = 2.5\mathrm{eV}$ for comparison with the calculation for zig-zag AFM order in fig. \ref{fig:Band_structure_U}(f). All three magnetic configurations are metallic but have considerably different band structures, producing different spectral functions and quasiparticle interference patterns. Fig. \ref{fig:Band_structure_U}(c)-(j) show the effect of increasing $U$ on the band structure of the zig-zag antiferromagnetic unit cell. While the inclusion of a $U$ term does change the band structure, its precise value does not have a significant effect, with variations resulting in only minor band shifts. However, for larger $U$ values the bands near the Fermi level have less Cr character. 

\subsection{Supporting ARPES data}
\begin{figure}
\centering
\includegraphics[scale=0.5]{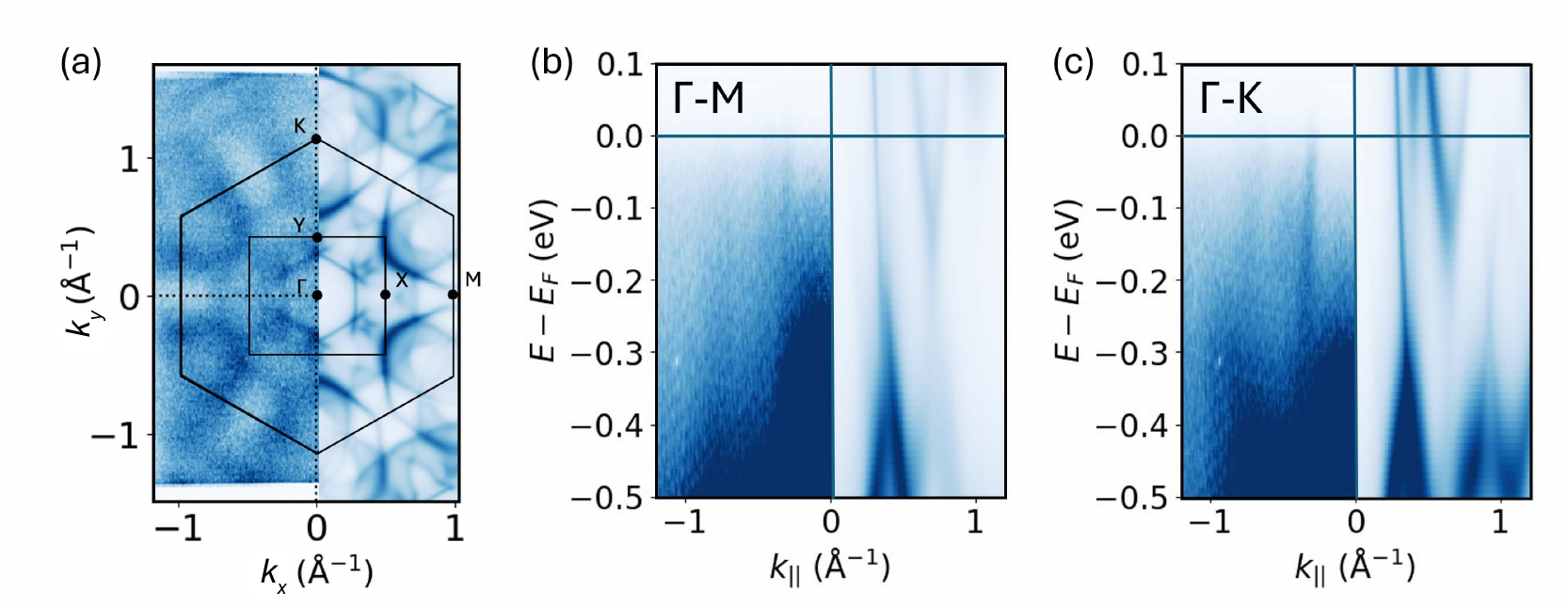}
\caption{ARPES measurements with photon energy $h\nu = 90 \mathrm{eV}$ and comparison with the symmetrized unfolded spectral function of the zig-zag antiferromagnetic order in \ce{CrTe2} with $U=2.5 \mathrm{eV}$. (a) Fermi surface from ARPES (left) and calculated spectral function (right), with $50\mathrm{meV}$ energy shift. (b) and (c) Cuts along (b) $\Gamma$-M and (c) $\Gamma$-K directions through ARPES data (left) and calculated spectral function (right), shifted by $50 \mathrm{meV}$.}
\label{fig:ARPES_90eV}
\end{figure}

Fig.~\ref{fig:ARPES_90eV} shows data acquired with a different photon energy, which selectively enhances the intensity of different Fermi pockets in our measurements. Irrespective of this, we note that the agreement with the calculations remains good, independent of the photon energy, as expected for a two-dimensional state.

\section{Comparison of DFT calculations to experiment}
\begin{figure}
\centering
\includegraphics[scale=0.45]{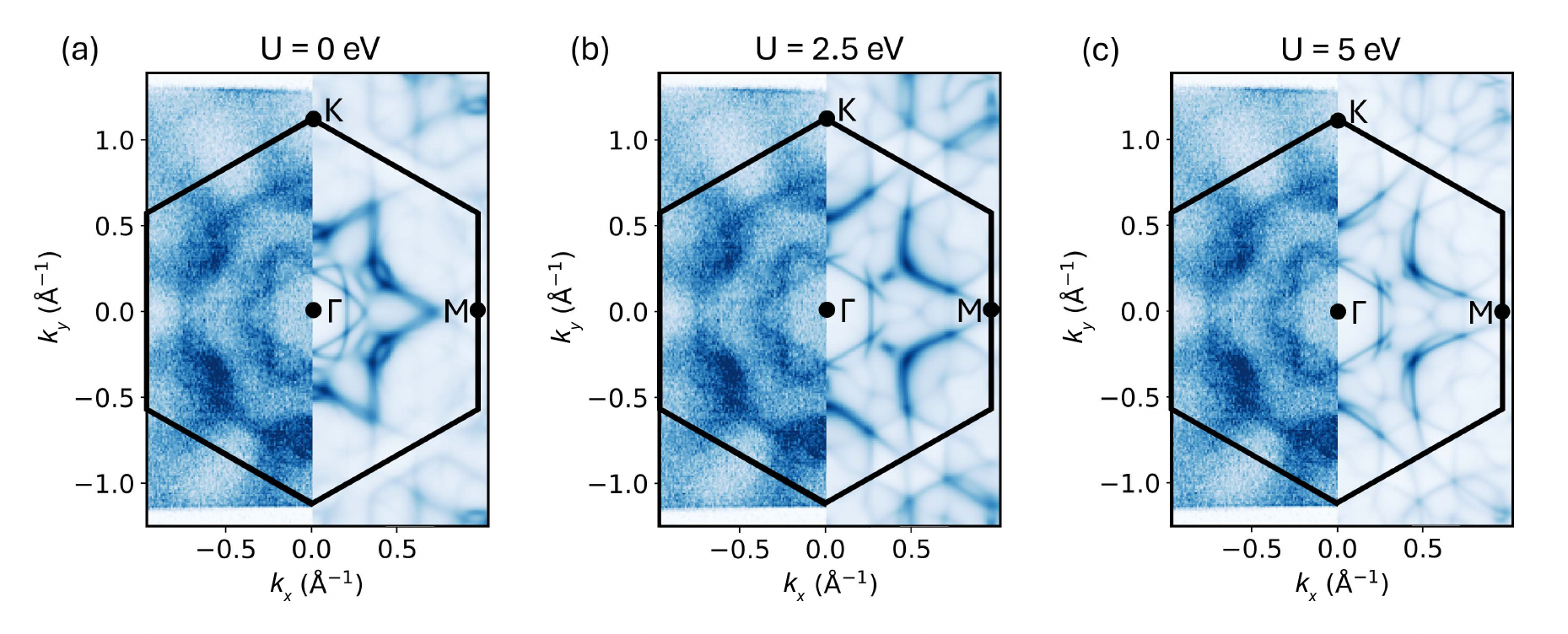}
\caption{Fermi surface as measured in ARPES (left) and calculated symmetrized unfolded spectral function calculations at $0 \mathrm{meV}$ of zigzag antiferromagnetic \ce{CrTe2} (right) with (a) $U=0\mathrm{eV}$, (b) $U=2.5\mathrm{eV}$ and (c) $U=5\mathrm{eV}$.}
\label{fig:ARPES_U_comparison}
\end{figure}
To determine the value of $U$ for which the DFT calculation best matches the experimental data, we have used the calculations shown in Fig.~\ref{fig:Band_structure_U} to obtain the unfolded spectral function (Fig. \ref{fig:ARPES_U_comparison}).   
Fig.~\ref{fig:ARPES_U_comparison} shows the effect of changing $U$ on the resulting symmetrized unfolded spectral function for $U=0\mathrm{eV}$, $2.5\mathrm{eV}$ and $5\mathrm{eV}$. From a comparison with the ARPES data, we conclude that a value of $2.5 \mathrm{eV}$ gives the best fit with experiment. 

\begin{figure}
\centering
\includegraphics[width=0.8\textwidth]{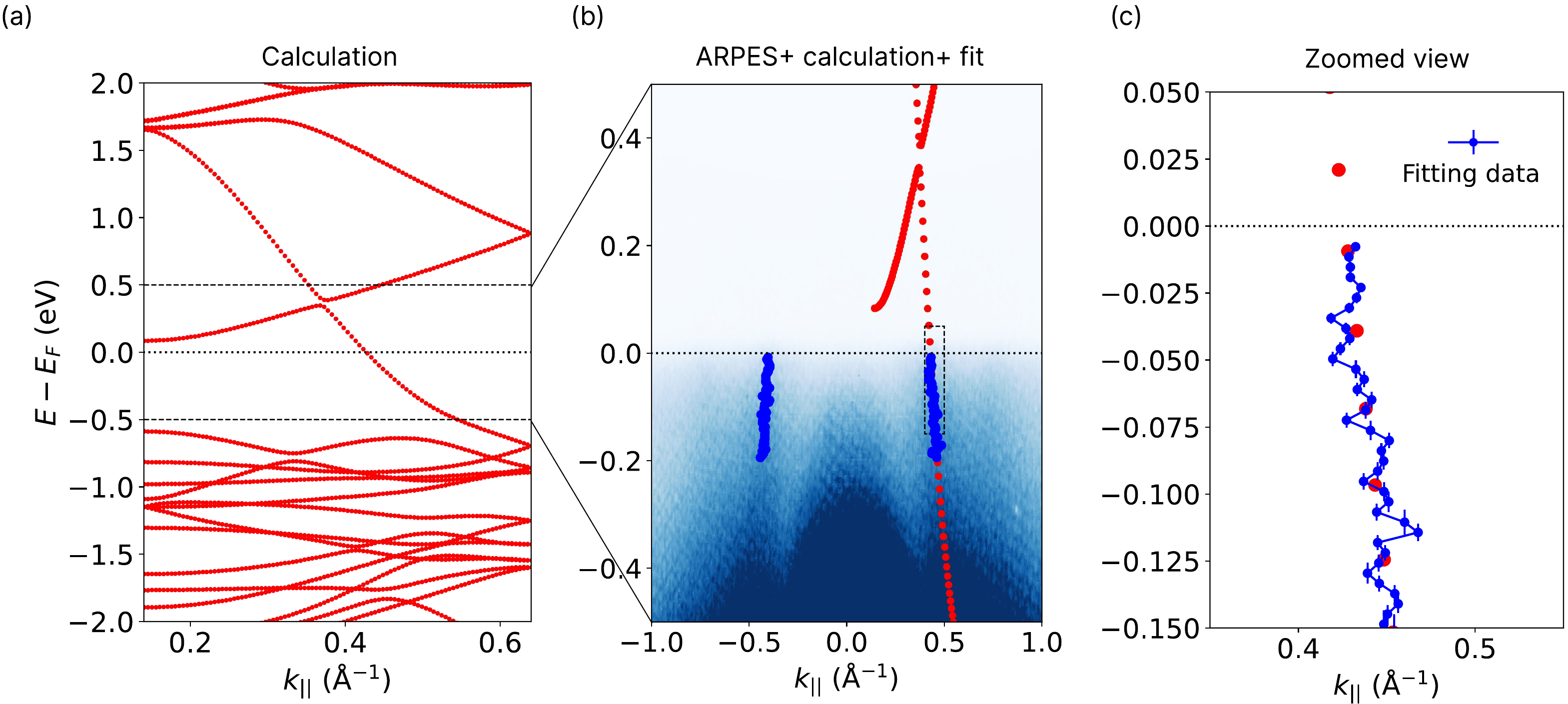}
\caption{Direct comparison of ARPES measurements and band structure from DFT, showing that no renormalization is required. (a) Band structure from the DFT calculation for a zig-zag antiferromagnetic unit cell with $U=2.5\mathrm{eV}$. (b) comparison of the ARPES data with the band structure from DFT using the same energy shift of $50\mathrm{meV}$ as used in the main text, and slightly shifting the DFT band structure by $0.14\mathrm{\AA}^{-1}$, so that the Fermi wavevectors match between the two. Fits to MDCs are shown as blue dots. (c) detail of (b) to highlight the agreement.}
\label{fig:ARPES_renorm}
\end{figure}
To evaluate whether there is evidence for an overall band renormalization from electronic correlations, we show in Fig.~\ref{fig:ARPES_renorm} a detailed comparison between the calculation and the ARPES data with a focus on whether the slope of the band is consistent. The comparison shows that, within the errors of the measurement, a good agreement is achieved without including any further renormalization in the comparison.

\begin{figure}
\centering
\includegraphics[scale=0.75]{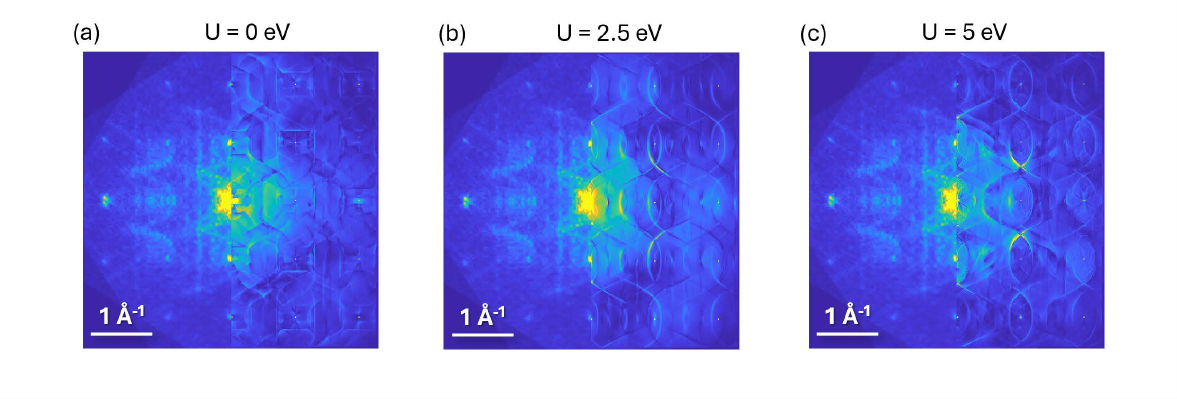}
\caption{Fourier transforms of $0 \mathrm{mV}$ STM map layer (left) compared with continuum local density of states calculations (with a positive scattering potential) at $0 \mathrm{meV}$ of zigzag antiferromagnetic \ce{CrTe2} (right) with (a) $U=0\mathrm{eV}$, (b) $U=2.5\mathrm{eV}$ and (c) $U=5\mathrm{eV}$.}
\label{fig:QPI_U_comparison}
\end{figure}
Fig.~\ref{fig:QPI_U_comparison} shows a comparison of the continuum local density of states calculation with different values of $U$ as in Fig.~\ref{fig:ARPES_U_comparison} with the measured QPI data. Again, a $U$ of $2.5 \mathrm{eV}$ gives the closest agreement with the data. 

\section{Influence of position of scatterer and scattering potential on QPI}
\begin{figure}
\centering
\includegraphics[scale=0.55]{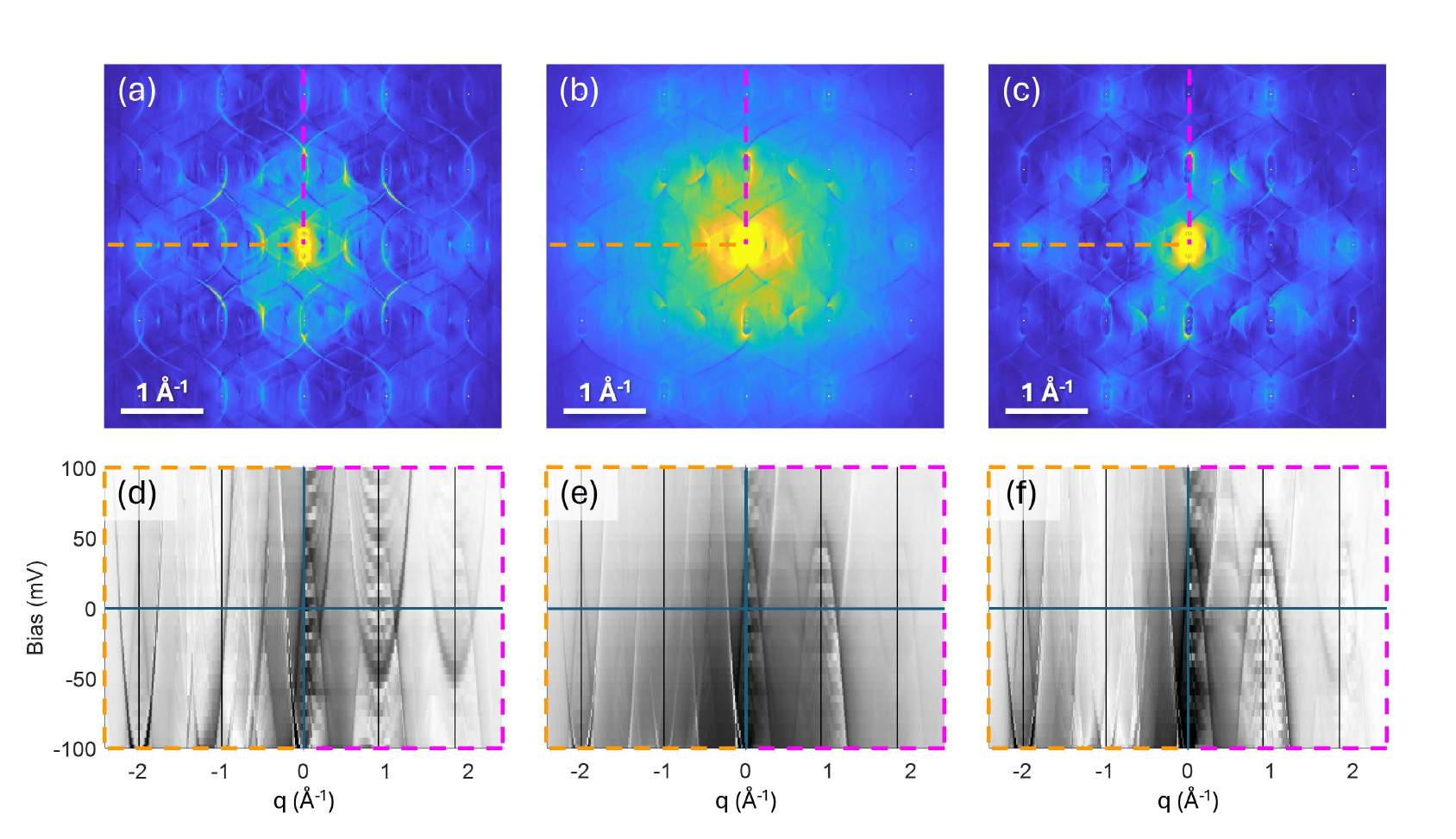}
\caption{Comparison of continuum local density of states calculations with scatterer on different atomic sites. All calculations have a positive scattering potential, a $U$ of $2.5\mathrm{eV}$ and include a $50 \mathrm{meV}$ energy shift. (a)-(c) Fourier transform at Fermi energy for (a) Cr, (b) surface Te and (c) sub-surface Te defects. (d)-(f) Cuts along high symmetry directions for (d) Cr, (e) surface Te and (f) sub-surface Te defects.}
\label{fig:Defect_comparison}
\end{figure}

\begin{figure}
\centering
\includegraphics[scale=0.55]{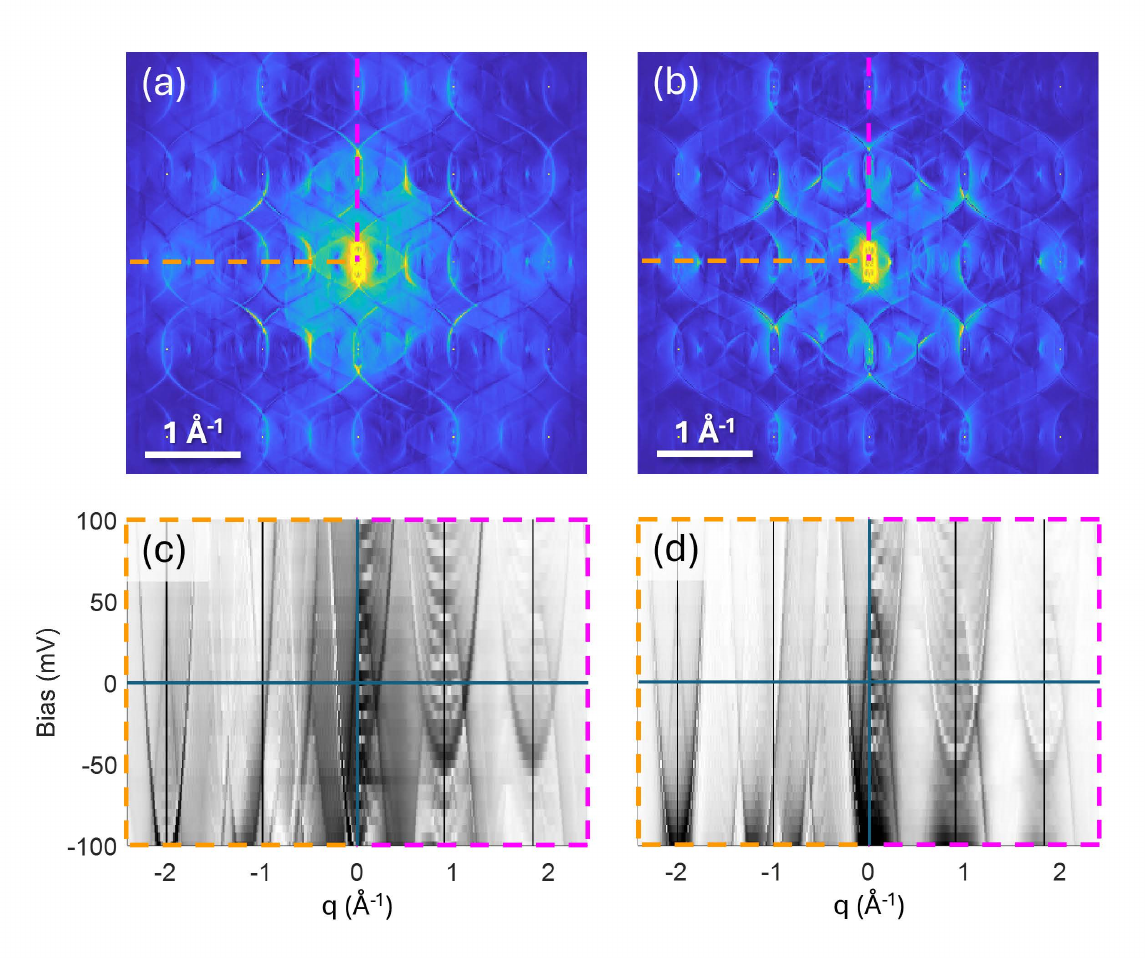}
\caption{Comparison of continuum local density of states calculations with different scattering potentials. Both calculations have a $U$ of $2.5\mathrm{eV}$ and include a $50 \mathrm{meV}$ energy shift. (a), (b) Fourier transforms at Fermi energy for (a) positive and (b) negative scattering potentials on a Cr site. (c), (d) Cuts along high symmetry directions for (c) positive and (d) negative scattering potentials.}
\label{fig:Potential_comparison}
\end{figure}

In Fig.~\ref{fig:Defect_comparison}, we show continuum local density of states calculations for scatterers at the three different sites: surface Te, sub-surface Te and Cr. All result in non-equivalent QPI patterns with differing intensities, however the same set of scattering vectors. While in reality in experiment, an ensemble of defects will result in a combination of the three contributions to the QPI signal, the best agreement with experiment is seen for Cr-site defects. Apart from the site of the defect, the specific potential can vary for defects of different chemical identity. In Fig.~\ref{fig:Potential_comparison}, we show two calculations performed with positive (a, c) and negative (b, d) scattering potentials. The calculations show again qualitatively the same features, but with different intensities.

\subsection{Influence of spin-orbit coupling on band structure}
\begin{figure}
\centering
\includegraphics[width=\textwidth]{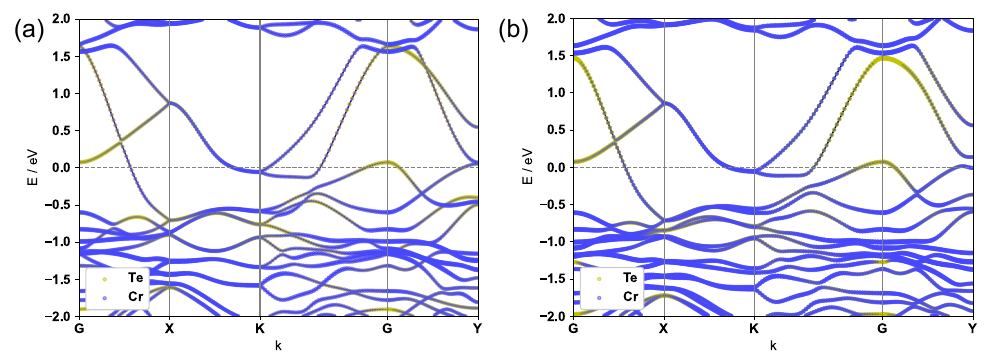}
\caption{Band structure obtained from DFT calculations with  $U=2.5\mathrm{eV}$ for the zig-zag antiferromagnetic order (a) without and (b) with spin-orbit coupling. While there are significant differences for some bands away from the Fermi energy, right at the Fermi energy the influence is very small.}
\label{fig:soc-comparison}
\end{figure}
In principle, Te exhibits a fairly large spin-orbit coupling, which might be expected to influence the band structure. For the bands close to the Fermi energy, this is surprisingly not the case, and the influence of spin-orbit coupling is rather negligible.
Fig. \ref{fig:soc-comparison}(a) and (b) show the band structures for the antiferromagnetic configuration and for $U = 2.5\mathrm{eV}$ with and without spin-orbit coupling. 

\section{Additional references}
\label{Bibliography}
\bibliographystyle{unsrtnat}
\bibliography{crte2}